\documentclass[prd,twocolumn,aps,amsmath,amssymb,nofootinbib,preprintnumbers]
{revtex4}
\usepackage[dvips]{graphicx}

\usepackage{graphicx}

\begin{document}



\newcommand{\be}{\begin{equation}}
\newcommand{\ee}{\end{equation}}
\newcommand{\ba}{\begin{eqnarray}}
\newcommand{\ea}{\end{eqnarray}}
\newcommand{\cH}{\cal{H}}


\title{Reheating in Inflationary Cosmology: Theory and Applications}


\author{Rouzbeh Allahverdi~$^{1}$,
Robert Brandenberger~$^{2}$,
Francis-Yan Cyr-Racine~$^{3}$,
Anupam Mazumdar~$^{4,~5}$}
\affiliation{$^{1}$~Physics Department, University of New Mexico,
Albuquerque, NM, 87131, USA \\
$^{2}$~Physics Department, McGill University, H3A 2T8, Canada. \\
$^{3}$~Department of Physics and Astronomy, Univ. of British Columbia,
Vancouver, BC, V6T 1Z1, Canada.\\
$^{4}$~Physics Department, Lancaster University, Lancaster LA1 4YB, United Kingdom.\\
$^{5}$~Niels Bohr Institute, Blegdamsvej-17, Copenhagen, DK-2100, Denmark.}


\begin{abstract}
Reheating is an important part of inflationary cosmology. It describes the production
of Standard Matter particles after the phase of accelerated expansion. We give
a review of the reheating process, focusing on an in-depth discussion of the
preheating stage which is characterized by exponential particle production due
to a parametric resonance or tachyonic instability. We give a brief overview of the thermalization
process after preheating and end with a survey of some
applications to supersymmetric theories and to
other issues in cosmology such as baryogenesis, dark matter and metric preheating.
\end{abstract}

\maketitle
\tableofcontents

\section{Introduction}

The inflationary model \cite{Guth} has become the current paradigm of early universe cosmology.
The first key aspect of the model is a phase of accelerated expansion of space which
can explain the overall homogeneity, spatial flatness and large size of the current
universe. Microscopic-scale quantum vacuum fluctuations during the phase of acceleration are
red-shifted to currently observable scales, and lead to a spectrum of cosmological
fluctuations which becomes scale-invariant in the limit in which the expansion rate
becomes constant in time \cite{ChibMukh}.

Reheating at the end of the period of accelerated expansion is an important
part of inflationary cosmology. Without reheating, inflation would leave behind
a universe empty of matter. Reheating occurs through coupling of the inflaton
field $\phi$, the scalar field generating the accelerated expansion of space,
to Standard Model (SM) matter. Such couplings must be present at least via
gravitational interactions. However, in many models of inflation there are
couplings through the matter sector of the theory directly.

Reheating was initially \cite{initial} analyzed using first order perturbation
theory and discussed in terms of the decay of an inflaton particle into
SM matter particles. As first realized in \cite{TB} (see also \cite{DK}),
such a perturbative analysis may be rather misleading since it does not
take into account the coherent nature of the inflaton field. A new view
of reheating was then proposed \cite{TB} which is based on the quantum
mechanical production of matter particles in a classical background inflaton
field \footnote{See also \cite{BHdV, Baacke} for other approaches to
the out-of-equilibrium dynamics of the inflaton field.}.
As this analysis showed, it is likely that reheating will involve
a parametric resonance instability. This proposal was studied more
carefully in \cite{KLS1,STB} and then analyzed in detail in \cite{KLS2}.
The term ``preheating" was coined \cite{KLS1} to describe the initial
energy transfer from the inflaton field to matter particles. Typically, the
state of matter after preheating is highly non-thermal, and thus must
be followed by a phase of thermalization.

The first goal of this review article is to present an introduction to the
theory of preheating after inflation. In Section 2 of this article, we
give a lightning review of inflationary cosmology. Section 3 is the
most important section of this review in which we present a comprehensive
analysis of preheating. The efficiency of preheating turns out to be
rather model-dependent. We first discuss ``standard preheating" which
will typically occur in simple single-field inflation models like those used
in ``Chaotic Inflation". The efficiency of preheating can be much higher
if a tachyonic direction develops, which is what occurs in certain small field
inflation models and in ``Hybrid Inflation", a model involving two scalar fields.
It turns out that while preheating leads to a very rapid start to the process
of energy transfer from the inflaton to SM matter, it typically
does not drain most of the energy of the inflaton field. This happens
in a second stage, a stage characterized by the nonlinear interactions
of the fluctuation modes which have been highly excited by the
preheating process. According to recent studies \cite{Micha}, this process is turbulent.
The initial stage of turbulence (after which the bulk of the energy density
is no longer in the inflaton field) is rapid, but the actual thermalization
of the decay products takes much longer. These issues are briefly
discussed in Section 4. Section 5 focuses on reheating in supersymmetric
models. Finally, in Section 6 we give a brief overview of a number of
applications of preheating in inflationary cosmology.


\section{Inflation Models and Initial Conditions for Reheating}

Cosmological inflation \cite{Guth} is a phase of accelerated expansion of space.
In the context of General Relativity as the theory describing space and time,
inflation requires scalar field matter. More precisely, the energy-momentum
tensor of matter must be dominated by the almost constant potential energy
density of the scalar field $\phi$.

A scalar field is postulated to exist in the SM of particle physics:
the Higgs field used to give elementary fermions their masses. To serve
as a Higgs field, its potential energy must have a minimum at a non-trivial
field value. The standard example is
\be \label{pot1}
V(\phi) \, = \, \frac{1}{4} \lambda (\phi^2 - \eta^2)^2
\ee
where $\eta$ is the vacuum expectation value of $\phi$. It is assumed that
at high temperatures the symmetry is restored by finite temperature
effects (see e.g. \cite{RHBRMP} for a review of field theory
methods used in inflationary cosmology) and $\phi = 0$. Once
the temperature $T$ falls below a critical value $T_c$, $\phi$ ceases
to be trapped and will start to roll towards one of the lowest energy
states $\phi = \pm \eta$. The SM Higgs
must have a coupling constant $\lambda$ which is set by the gauge
coupling constant and cannot be sufficiently small to yield a long
time period of slow rolling of $\phi$ which is required to obtain enough
inflation (except possibly if $\phi$ is non-minimally coupled to gravity
\cite{Shaposh}).

Hence, scalar field-driven inflation requires us to go beyond the
SM of particle physics. Once one makes this step, there
are typically many candidate scalar fields which could be the inflaton,
in particular in supersymmetric models. 

For cosmological studies, the precise nature of the inflaton is often 
secondary and hence simple toy models are used.
``New" inflation \cite{Linde1,AS} maintains the idea that $\phi$ begins
trapped near $\phi = 0$. Inflation takes place during the period
when $\phi$ is undergoing the symmetry-breaking phase transition
and slowly rolling towards $\phi = \pm \eta$. The model was based
on scalar field dynamics obtained by replacing the potential (\ref{pot1})
by a symmetry breaking potential
of Coleman-Weinberg \cite{CW} form, where the mass term at the
field origin is set to zero and symmetry breaking is obtained through
quantum corrections. However, new inflation models typically suffer
from an initial condition problem \cite{Piran}.

``Chaotic" (or ``large-field") inflation \cite{Linde2} is an alternative
scenario. Inflation is triggered by a period of slow-rolling of $\phi$
unrelated to a symmetry-breaking phase transition. The
simplest example occurs in the toy model of a single scalar field
with potential
\be \label{pot2}
V(\phi) \, = \, \frac{1}{2} m^2 \phi^2 \,
\ee
where $m$ is the mass of $\phi$ (which is of the order $10^{-6} m_{pl}$
if the model is to yield the
observed magnitude of cosmological fluctuations \cite{ChibMukh}).
Here, $m_{pl}$ is the Planck mass defined via $m_{pl}^{-2} \equiv G$,
$G$ being Newton's gravitational constant. It is assumed that  $\phi$ starts
out at large field values and slowly rolls towards its vacuum
state $\phi = 0$. For inflation to be successful, two conditions
must be satisfied. Firstly, the energy density must be dominated
by the potential energy term, and secondly the acceleration
term in the field equation
\be \label{eom}
{\ddot{\phi}} + 3 H {\dot{\phi}} \, = \, - V^{'}(\phi) \, ,
\ee
the Klein-Gordon equation in an expanding background, must be
negligible compared to the other two terms. Here, $H$ is the
Hubble expansion rate and a prime indicates the derivative
with respect to $\phi$. Making use of the Friedmann equation
\be \label{FRW}
H^2 \, = \, \frac{8 \pi G}{3} \left( \frac{1}{2} {\dot{\phi}}^2 + V(\phi) \right)
\ee
it is easy to see that the slow-rolling conditions are only satisfied
for super-Planckian field values $|\phi| > m_{pl}$.
In the above two equations (\ref{eom}) and (\ref{FRW}) we have
taken the field configuration to be homogeneous. In the case of
chaotic inflation, the homogeneous slow-roll trajectory is
a local attractor in initial condition space \cite{Kung}, even in
the presence of linear metric fluctuations \cite{Hume}, and thus
this model is free from the initial condition problem of new inflation. 
In the context of ``real" particle physics theories such as
supersymmetric models,  gravitational effects often
steepen the potential for values of $|\phi|$ beyond the
Planck mass and therefore prevent slow-roll inflation. 

One way to try to avoid this problem but maintain the success of 
chaotic inflation is to add a second scalar field $\psi$ to the sector of the theory
responsible for inflation and to invoke a potential of the form
\be \label{hybrid}
V(\phi, \psi) \, = \, \frac{1}{2} m^2 \phi^2 + \frac{g^2}{2} \psi^2 \phi^2
+ \frac{1}{4} \lambda \bigl( \psi^2 - v^2 \bigr)^2 \, ,
\ee
where $g$ and $\lambda$ are dimensionless coupling constants
and $v$ is the vacuum expectation value of $\psi$.
For large values of $|\phi|$, the potential in $\psi$
direction has a minimum
at $\psi = 0$, whereas for small values of $|\phi|$, $\psi = 0$
becomes an unstable point.
The reader can verify that in this model slow-rolling
of $\phi$ does not require super-Planckian field values. This two
field model is called ``hybrid" inflation \cite{Linde3}.

Let us return to the toy model of chaotic inflation with the potential
(\ref{pot1}). The slow-roll trajectory is given by
\be
{\dot{\phi}} \, = \, - \frac{1}{2 \sqrt{3} \pi} m m_{pl} \, ,
\ee
and it is easy to see that the slow-roll conditions break down at the
field value
\be
\phi_c \ = \, \frac{m_{pl}}{2 \sqrt{3 \pi}} \, .
\ee
After the breakdown of slow-rolling, $\phi$ commences damped
oscillatory motion about $\phi = 0$ and the time-averaged equation
of state is that of cold matter ($p = 0$ where $p$ denotes pressure).
Asymptotically for large times $m t \gg 1$ the solution approaches
\be
\phi(t) \, \rightarrow \, \frac{m_{pl}}{\sqrt{3 \pi} mt} {\rm sin}(mt) \, .
\ee
This scalar field configuration will provide the classical background
matter in the reheating phase.


\section{Inflaton Decay}\label{section_reheating}

\subsection{Perturbative Decay}

Reheating is a key part of inflationary cosmology. It describes the
production of SM matter at the end of the period
of accelerated expansion when the energy density is stored
overwhelmingly in the oscillations of $\phi$.  Historically, reheating
was first treated perturbatively \cite{initial}.

We assume that the inflaton $\phi$
is coupled to another scalar field $\chi$. 
Taking the interaction Lagrangian to be
\be \label{toy}
{\cal L}_{\rm int} \, = \, - g \sigma \phi \chi^2 \, ,
\ee
where $g$ is a dimensionless coupling constant and $\sigma$
is a mass scale, then the decay rate of the inflaton into
$\chi$ particles is given by
\be
\Gamma \, = \, \frac{g^2 \sigma^2}{8 \pi m} \, ,
\ee
where $m$ is the inflaton mass.

In the approach of \cite{initial}, the energy loss of the inflaton due
to the production of $\chi$ particles was taken into account by
adding a damping term to the inflaton equation of motion
which in the case of a homogeneous inflaton field is
\be \label{effeom}
{\ddot \phi} + 3 H {\dot \phi} + \Gamma {\dot \phi} \, = \, - V^{'}(\phi) \, .
\ee
For small coupling constant, the interaction rate $\Gamma$ is
typically much smaller than the Hubble parameter at the end
of inflation. Thus, at the beginning of the phase of inflaton
oscillations, the energy loss into particles is initially negligible
compared to the energy loss due to the expansion of space.
It is only once the Hubble expansion rate decreases to a
value comparable to $\Gamma$ that $\chi$ particle production
becomes effective. It is the energy density at the time when
$H = \Gamma$ which determines how much energy ends up
in $\chi$ particles and thus determines the ``reheating temperature",
the temperature of the SM fields after energy transfer.
\be
T_R \, \sim \, \left( \Gamma m_{pl} \right)^{1/2} \, .
\ee
Since $\Gamma$ is proportional to the square of the coupling
constant $g$ which is generally very small, perturbative
reheating is slow and produces a reheating temperature
which can be very low compared to the energy scale
at which inflation takes place.

There are two main problems with the perturbative decay
analysis described above. First of all, even if the inflaton
decay were perturbative, it is not justified to use the heuristic
equation (\ref{effeom}) since it violates the fluctuation-dissipation
theorem: in systems with dissipation, there are always fluctuations,
and these are missing in (\ref{effeom}). For an improved
effective equation of motion see e.g. \cite{Ramos}.

The main problem with the perturbative analysis is that it does
not take into account the coherent nature of the inflaton field.
The inflaton field at the beginning of the period of oscillations
is not a superposition of free asymptotic single inflaton states,
but rather a coherently oscillating homogeneous field. The large
amplitude of oscillation implies that it is well justified to treat
the inflaton classically. However, the 
matter fields can be assumed to start off in their vacuum state
(the red-shifting during the period of inflation will remove any
matter particles present at the beginning of inflation). Thus,
matter fields $\chi$ must be treated quantum mechanically.
The improved approach to reheating initiated in \cite{TB}
(see also \cite{DK}) is to consider reheating as a quantum
production of $\chi$ particles in a classical $\phi$ background.

\subsection{Preheating}

We will present the preheating mechanism for the simple
toy model with interaction Lagrangian
\be \label{intLag}
{\cal L}_{\rm int} \, = \, - \frac{1}{2} g^2 \chi^2  \phi^2 \, ,
\ee
where, as before, $g$ is a dimensionless coupling constant
\footnote{Preheating in a conformally flat scalar field model
was analyzed in \cite{Kaiser, GKLS}, and in a sine-Gordon potential
in \cite{GKS}.}.
In this subsection we will neglect the expansion of space.
Provided that the time period of preheating is small compared
to the Hubble expansion time $H^{-1}$ this is a reasonable
approximation. In the next subsection we will include the
expansion of space explicitly.

The quantum theory of $\chi$ particle production in the external classical
inflaton background begins by expanding the quantum field $\hat{\chi}$ into
creation and annihilation operators $\hat{a}_{k}$ and $\hat{a}^{\dagger}_{k}$ as:
\be
\hat{\chi}(t,\mathbf{x})=\frac{1}{(2\pi)^{3/2}}\int d^3k \left(\chi_k^*(t)\hat{a}_{k} e^{i\mathbf{k}\mathbf{x}}+\chi_k(t)\hat{a}^{\dagger}_{k} e^{-i\mathbf{k}\mathbf{x}}\right),
\ee
where $k$ is the momentum. If we assume that there are no non-linearities in the $\chi$ sector
of the theory, then the equation of motion for $\chi$ is linear and
can be studied simply mode by mode in Fourier space. The mode functions then satisfy
the equation
\be\label{math3}
\ddot{\chi}_k+\left(k^2+m_{\chi}^2+g^2\Phi^2\sin^2{(mt)}\right)\chi_k \, = \, 0 \, ,
\ee
where $\Phi$ is the amplitude of oscillation of $\phi$.
This is the Mathieu equation which is conventionally written in
the form
\be\label{mat}
\chi''_k+(A_k-2q\cos{2z})\chi_k \, = \, 0 \, ,
\ee
where we have introduced the dimensionless time variable $z = m t$
and a prime now denotes the derivative with respect to $z$.
Comparing the coefficients, we see that
\be\label{param1}
A_k=\frac{k^2+m_{\chi}^2}{m^2}+2q\qquad q=\frac{g^2\Phi^2}{4m^2}
\ee
The growth of the mode function
corresponds to particle production, as in the case of particle production in
an external gravitational field \cite{Birrell}. We will return to this point in the next
subsection. For now, let us simply state that exponential growth of the mode
functions will lead to an exponential growth of the number of $\chi$ particles,
with the exponent of this growth being twice the corresponding exponent of the
mode functions.

It is well known that the Mathieu equation has instabilities for certain ranges of $k$
and leads to exponential growth
\be
\chi_k \, \propto \, {\rm exp}(\mu_k z) \, ,
\ee
where $\mu_k$ is called the Floquet exponent. For small values of $q$, e.g. $q \ll 1$,
resonance occurs in a narrow instability band about $k = m$ (see Figure 1).
Hence, in this case we speak
of ``narrow resonance" (see \cite{Mathieu} for in-depth discussions of the Mathieu
equation and its generalizations). 

\begin{figure}
\includegraphics[height=8cm]{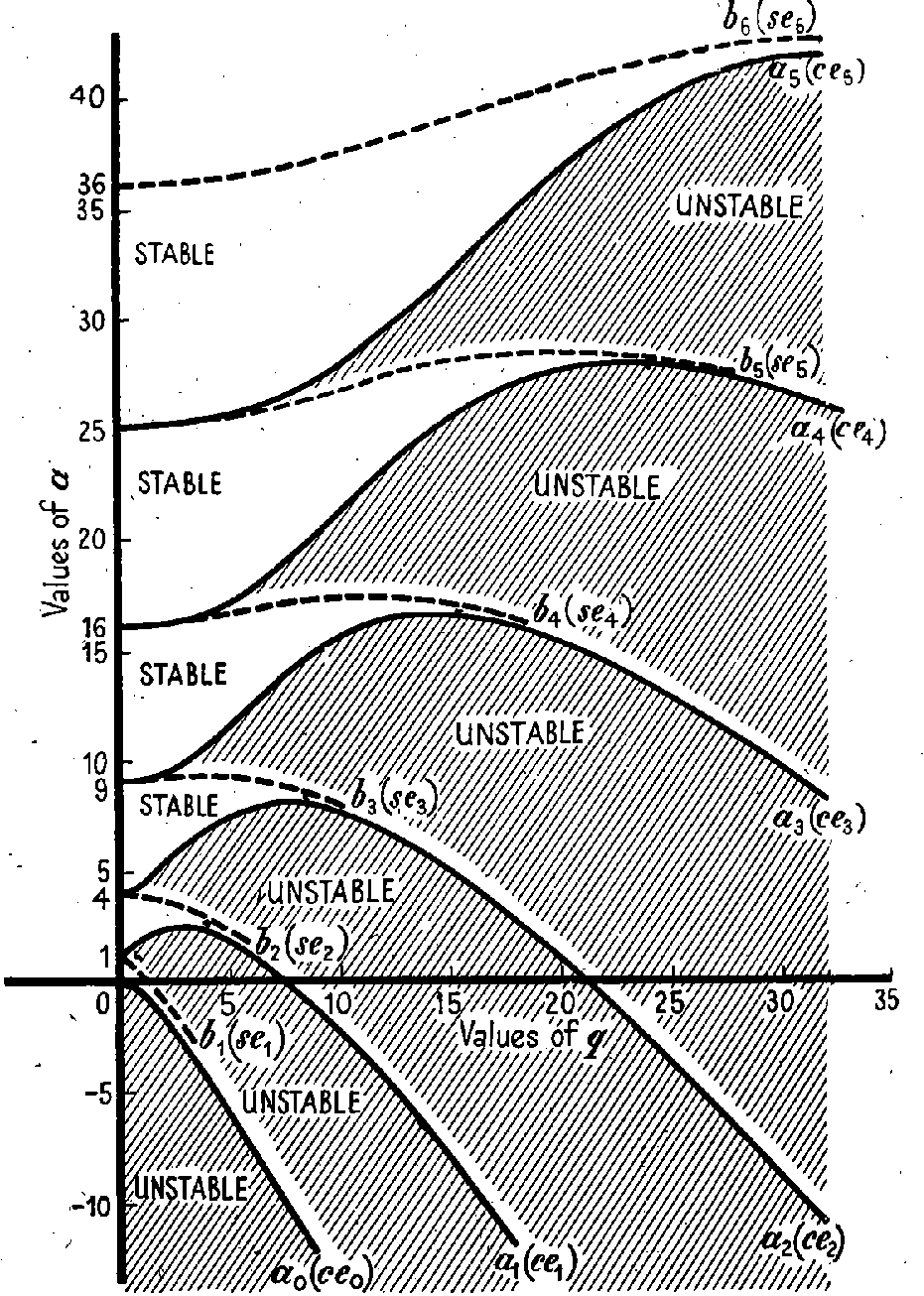}
\caption{Instability bands of the Mathieu equation (from \cite{Mathieu}). The horizontal
axis is the parameter $q$ of (\ref{mat}), the vertical axis is the value of $A$. The shaded
regions are regions in parameter space where there is a parametric resonance instability.}
\label{fig1}
\end{figure}

The resonance is much more efficient if $q \gg 1$ \cite{KLS1,KLS2}.
In this case, resonance occurs in broad bands. In particular, the bands include all
long wavelength modes $k \rightarrow 0$. We then speak of ``broad" parametric
resonance. A condition for particle production is that the WKB approximation for
the evolution of $\chi$ is violated. In the WKB approximation, we write:
$\chi_k \, \propto \, e^{\pm i\int\omega_kdt}$,
which is valid as long as the adiabaticity condition
\be \label{adiab}
\frac{d\omega_k^2}{dt} \, \leq \, 2 \omega^3_k
\ee
is satisfied. In the above, the effective frequency $\omega_k$ is given by
\be \label{efffreq}
\omega_k \, = \, \sqrt{k^2 + m_{\chi}^2 + g^2 \Phi(t)^2 sin^2(mt)} \, ,
\ee
By inserting the effective frequency (\ref{efffreq}) into the condition (\ref{adiab}) and
following some algebra, we find that the adiabaticity condition is violated for momenta satisfying
\be
k^2 \, \leq \, \frac{2}{3\sqrt{3}}gm\Phi-m_{\chi}^2.
\ee
For modes with these values of $k$, the adiabaticity condition breaks
down in each oscillation period when $\phi$ is close to zero. We
conclude that the particle number does not increase smoothly, but
rather in ``bursts", as was first studied in \cite{KLS2}.

So far, we have studied preheating in a toy model in which Standard
Model matter is modeled by a scalar field $\chi$. However, in
principle we are interested in the production of SM fermions.
Such fermions could be produced after preheating into a scalar field
$\chi$ which then in turn couples to fermions. In particular, in supersymmetric
theories to be discussed in a later section there are many channels for
this to happen. However, it turns out that preheating into fermions is
also effective, in spite of the fact that the occupation number of any fixed state
cannot be greater than one (because of the
Pauli exclusion principle). This is discussed in detail in
\cite{fermionic,fermionic2}. 

\subsection{Preheating in an Expanding Background}

Provided that the Floquet exponent is not much smaller than unity,
the parametric resonance instability leads to an energy transfer
from the inflaton to matter particles which is rapid on the scale of
the Hubble time. Thus, an analysis neglecting the expansion of
the universe is self-consistent. However, as discussed in detail
in \cite{KLS2}, it is not too difficult to include the expansion of
space.

We will first give a qualitative analysis of broad resonance in
an expanding background characterized by the cosmological
scale factor $a(t)$ which is increasing in time as given by the
Friedmann equations. The equation of motion for $\chi$ is
\be\label{eom_chi_frw}
\ddot{\chi}_k+3H\dot{\chi}_k+\left(\frac{k^2}{a^2}+m_{\chi}^2+
g^2\Phi(t)^2\sin^2{(mt)}\right)\chi_k \, = \, 0.
\ee
The adiabaticity condition is now violated for momenta satisfying:
\be\label{ineq_adia2}
\frac{k^2}{a^2} \, \leq \, \frac{2}{3\sqrt{3}}gm\Phi(t)-m_{\chi}^2.
\ee
Note that the expansion of space makes broad resonance more
effective since more $k$ modes are red-shifted into
the instability band as time proceeds. We will see below that the
improved analysis yields the same expression for the resonance
band except for the exact value of the numerical coefficient of the
first term on the r.h.s..
Broad parametric resonance ends when $q \leq 1/4$.

Let us now move on to a quantitative analysis of this problem, building on the
comprehensive study of \cite{KLS2}. It proves convenient to eliminate the
Hubble friction term in the equation of motion by rescaling the field variable. We
consider the variable $X_k(t)=a^{3/2}(t)\chi_k(t)$ in terms of which
the equation of motion (\ref{eom_chi_frw}) becomes:
\be\label{eom_X}
\ddot{X}_k+\omega_k^2X_k \,= \, 0
\ee
with
\be\label{exp_freq}
\omega_k^2 \, = \, \frac{k^2}{a^2(t)}+m_{\chi}^2+g^2\Phi^2(t)\sin^2{(mt)}
-\frac{9}{4}H^2-\frac{3}{2}\dot{H}.\ee
Note that in the matter-dominated background which we are considering the
last two terms on the right-hand side cancel.

The equation of motion (\ref{eom_X}) represents a harmonic oscillator equation
with a time-dependent frequency. The evolution of the solution will be described
by the WKB approximation (which entails the absence of particle
production) unless the adiabaticity condition is violated. This will happen during
short time intervals around the instances $t = t_j$ when $\phi = 0$, as discussed in
the previous section. We label the intervals of adiabatic evolution by an integer
$j$. In the j'th interval (lasting from $t_{j - 1}$ to $t_j$),
the adiabatic evolution of $X_k$ is given by
\be\label{full_sol_X}
X_k^j(t) \, = \, \frac{\alpha_k^j}{\sqrt{2\omega_k}}e^{i\int\omega_kdt}+\frac{\beta_k^j}{\sqrt{2\omega_k}}e^{-i\int\omega_kdt},
\ee
where the coefficients $\alpha_k^j$ and $\beta_k^j$ (the ``Bogoliubov coefficients'')
are constant and satisfy the normalization condition $|\alpha_k^j|^2-|\beta_k^j|^2=1$ 
(derived from the Heisenberg uncertainty principle). 

During the brief time periods when $\phi$ is close to zero, we can use the
approximation $\phi^2(t) \simeq \Phi^2 m^2 (t - t_j)^2$. Introducing the
 new time variable $\tau = g \Phi m (t - t_j)$ and a rescaled momentum
 $\kappa^2 = \frac{(k^2/a^2) + m_{\chi}^2}{g \Phi m}$ the equation of motion
 for $X$ becomes
 \be
 \frac{d^2 X_k}{d \tau^2} + \bigl( \kappa^2 + \tau^2 \bigr) X_k \, = \, 0 \, .
 \ee
 This equation corresponds to scattering from a parabolic potential.

The non-adabatic evolution of $X_k$ during the short intervals when $\phi$ crosses
the origin leads to a transformation of the coefficients of Bogoliubov type
\ba\label{bogo_transf}
\left(\begin{array}{c}\alpha_k^{j+1}\\ \beta_k^{j+1}\end{array}\right)&=&\left(\begin{array}{cc}\frac{1}{D_k}&\frac{R^*_k}{D^*_k}e^{-2i\theta_k^j}\\\frac{R_k}{D_k}e^{2i\theta_k^j}&\frac{1}{D^*_k}\end{array}\right)\left(\begin{array}{c}\alpha_k^{j}\\ \beta_k^{j}\end{array}\right)\ea\ba&=&\left(\begin{array}{cc}\sqrt{1+e^{-\pi\kappa^2}}e^{i\varphi_k}&ie^{-\frac{\pi}{2}\kappa^2-2i\theta_k^j}\\-ie^{-\frac{\pi}{2}\kappa^2+2i\theta_k^j}&\sqrt{1+e^{-\pi\kappa^2}}e^{-i\varphi_k}\end{array}\right)\left(\begin{array}{c}\alpha_k^{j}\\ \beta_k^{j}\end{array}\right), \nonumber
\ea
where we defined $\theta_k^j=\int_0^{t_j}dt\omega_k$ which is the phase accumulated at $t_j$.
The reflection and transmission coefficients $R_k$ and $D_k$ are given by
\ba
R_k \, & = \, &-\frac{ie^{i\varphi_k}}{\sqrt{1+e^{\pi\kappa^2}}}\\
D_k \, & = \, & \frac{e^{-i\varphi_k}}{\sqrt{1+e^{-\pi\kappa^2}}},
\ea
where the phase $\varphi_k$ is:
\be
\varphi_k \, = \, \rm{arg}\left\{\Gamma\left(\frac{1+i\kappa^2}{2}\right)\right\}+\frac{\kappa^2}{2}\left(1+\ln{\frac{2}{\kappa^2}}\right) \, .
\ee
Here $\Gamma$ stands for the complex Gamma function. 
In accordance with the general theory of particle production in external fields (see e.g. \cite{Birrell}),
the occupation number of the $k$'th mode is
\be\label{occupation_number1}
n_k \, = \, |\beta_k|^2.
\ee
Making use of the Bogoliubov transformation (\ref{bogo_transf}), we obtain the following
recursion relation of the particle number:
\ba\label{occ_num_k}
n_k^{j+1} \, &=& \, e^{-\pi\kappa^2}+\left(1+2e^{-\pi\kappa^2}\right)n_k^j \nonumber \\
&+& \, 2e^{-\frac{\pi}{2}\kappa^2}\sqrt{1+e^{-\pi\kappa^2}}\sqrt{n_k^j(1+n_k^j)}\sin{\theta_{tot}^j},
\ea
where $\theta_{tot}^j=2\theta_k^j-\varphi_k+\rm{arg}(\alpha_k^j)-\rm{arg}(\beta_k^j)$, we used (\ref{occupation_number1}) and  $|\alpha_k^j|^2=1+n_k^j$. From (\ref{occ_num_k})
it follows that only modes with $\pi \kappa^2 \leq 1$ will grow. Inserting the definition
of $\kappa$ into this condition, we get an expression for the resonance band
which reproduces (\ref{ineq_adia2}) except for the numerical coefficient in the first
term on the r.h.s. of the equation which is now $1/\pi$ instead of $2/(3 \sqrt{3})$.

We can expand (\ref{occ_num_k}) in the limit $n_k\gg1$ and obtain
\be\label{occ_num_large}
n_k^{j+1} \, = \, \left(1+2e^{-\pi\kappa^2}
+2e^{-\frac{\pi}{2}\kappa^2}\sqrt{1+e^{-\pi\kappa^2}}\sin{\theta_{tot}^j}\right)n_k^j.
\ee
The occupation number can only grow if the expression in the bracket is larger than $1$. For the fastest growing mode ($k=0$), $n_k^{j+1} > n_k^j$ if
\be\label{range_theta}
-\frac{\pi}{4}\, < \, \theta_{tot}^j \, < \, \frac{5\pi}{4} \, ,
\ee
where we have taken $m_{\chi}=0$. Now, since $\theta_{tot}^j$ has a rather complicated time-dependence, it could almost be considered as a random variable. Thus, the range of phases (\ref{range_theta}) implies that the solution grows about $75\%$ of the time. 

We can define an effective Floquet exponent $\mu_k^j$ for the j'th interval via
\be
n_k^{j + 1} \, = \, e^{2 \pi \mu_k^j} n_k^j \, .
\ee
By comparing with (\ref{occ_num_large}) we get
\be\label{floquet}
\mu_k^j \, = \, \frac{1}{2\pi}\ln{\left(1+2e^{-\pi\kappa^2}
+2e^{-\frac{\pi}{2}\kappa^2}\sqrt{1+e^{-\pi\kappa^2}}\sin{\theta_{tot}^j}\right)}.
\ee
The ``average'' Floquet exponent $\mu_k$ is determined by:
\be\label{mu_k}
\mu_k \ = \, \frac{\pi}{m\Delta t}\sum_j\mu_k^j \, ,
\ee
where $\Delta t$ is the total duration of the resonance. 
Making use of (\ref{floquet}) we obtain
\be\label{scale_dep_floquet}
\mu_k \, \approx \, \frac{1}{2\pi }\ln{3} \, - \, \mathcal{O}(\kappa^2)\, . 
\ee
The fact that the exponent is of the order unity implies that broad parametric resonance
in an expanding background is very efficient and can convert a substantial fraction 
of the inflaton energy density into matter in a time interval small compared to the Hubble time.

The total number density of $\chi$ particles is obtained by integrating over all values of $k$:
\ba\label{integral_n_chi}
n_{\chi}(t) \, &=& \, \frac{1}{(2\pi a)^3}\int_0^{\infty} d^3kn_k(t) \\
&=& \, \frac{1}{2\pi^2a^3}\int_0^{\infty} dkk^2|\beta_k^0|^2e^{2m\mu_kt}. \nonumber
\ea
which can be estimated as \cite{FrancisThesis}
\be\label{estimate_nchi}
n_{\chi}(t)\approx\frac{(gm\Phi_0)^{3/2}|\beta^0_m|^2e^{2m\mu t}}{16\pi^3a^3\sqrt{\frac{m\mu t}{2\pi}+1}}.
\ee
This equation determines how fast the energy is drained from the inflaton field, and thus the
time interval which it takes before preheating is completed.

\subsection{Termination of Preheating}

In the previous analysis of preheating, we have neglected the back-reaction of the
produced $\chi$ particles on the dynamics of the preheating process. The back-reaction
arises at different places. First, the presence of $\chi$ particles changes the effective
mass of the inflaton oscillations. The rough criterion which we will use below is that
this back-reaction effect is negligible as long as the change $\Delta m_{\phi}^2$ in the
square mass of the inflaton is smaller than $m^2$.

In the Hartree approximation, the change in the inflaton mass due to $\chi$ particles is given by
\be
\Delta m_{\phi}^2 \, = \, g^2 \langle \chi^2 \rangle \, ,
\ee
where the pointed brackets indicate the quantum expectation value.
The expectation value of $\chi^2$ is given by
\be
\label{hartree_app}
\langle\chi^2\rangle \, = \, \frac{1}{2\pi^2 a^3}\int_0^{\infty}dk k^2|X_k(t)|^2
\ee
Inserting the expansion of the field modes $X_k$ in terms of the Bogoliubov coefficients we
find that
\be \label{effmass}
\Delta m_{\chi}^2(t) \, \simeq \, \frac{g n_{\chi}(t)}{|\phi(t)|} \, .
\ee
It appears that this expression becomes ill-defined when $\phi$ crosses zero. However,
the equation of motion is still well defined at these points. To estimate the
strength of back-reaction, we will replace $\phi$ by its amplitude $\Phi$. Thus, the
condition under which this back-reaction effect is negligible is
\be
n_{\chi}(t) \, \leq \, \frac{m^2 \Phi(t)}{g} \, .
\ee
This is an implicit equation for the time $t_1$ when back reaction can no longer
be neglected. Note that the expression for the number density of $\chi$ particles
was derived in (\ref{estimate_nchi}), and that $\Phi$ scales as a function of time
as $t^{-1}$. Because of the exponential growth of $n_{\chi}$, it is clear that
up to logarithmic factors the time interval of preheating is
\be
\delta t \, \sim \, (\mu m)^{-1} \, .
\ee
Thus, since $H \ll m$, this time interval is short compared to the Hubble expansion
time, unless the Floquet exponent $\mu$ is suppressed.

A second condition must be satisfied in order to be able to neglect the back-reaction
of the $\chi$ particles on the preheating dynamics: it is the condition that the
energy in the $\chi$ particles is sub-dominant. Making us of the estimate,
$\rho_{\chi} \, \sim \, \langle (\nabla \chi)^2 \rangle \, \simeq \, k^2 \langle \chi^2 \rangle $,
and inserting the value of $\langle \chi^2 \rangle$ at the time $t_1$ determined above,
we see that $\rho_{\chi}$ is smaller than the potential energy of the inflaton
field at the time $t_1$ as long as the value $q$ at the time $t_1$ is larger
than $1$, e.g. $q(t_1) > 1$. This is roughly speaking the same as the
condition for the effectiveness of broad resonance.

As it turns out, in many models there is another mechanism which shuts
off the resonance before the either of the two conditions mentioned above
becomes satisfied. Numerical studies \cite{KT1,KT2,KT3,Tom} have shown that the
scattering of $\chi$ particles off the inflaton condensate limits the
value of the $\chi$ modes to a value lower than that which would be
obtained from arguments such as the first one above which makes use
of the Hartree approximation. For small values of $q$, the resonant
period may be completely absent.

If $q(t_1) >> 1$ then broad parametric resonance ends when most of the
energy is still stored in the inflaton condensate. The further decay must
then be analyzed by other techniques, e.g. perturbatively or using
numerical simulations. In particular,
the time interval of matter-dominated dynamics after inflation could last
a long time, leading to a low matter temperature after the matter field
excitations have thermalized. On the other hand, if $q(t_1) \sim 1$,
then the matter-dominated phase will end with the end of preheating.
We still need to study how long it takes for the decay products to
thermalize. This topic of thermalization after preheating will be
discussed in the next section.

Before moving on to the study of thermalization after preheating,
we must discuss some variants of preheating in which the
inflaton decay is much more efficient than in the chaotic inflation
toy model used so far in this section.

\subsection{Tachyonic Preheating}

In the chaotic inflation model we have studied up to this point
the effective frequency of the $\chi$ oscillations is always
positive. If it were negative, then we would obviously get
an exponential instability. The simplest way to obtain
this instability is to simply change the sign of the
coupling term (\ref{intLag}) in the interaction Lagrangian. This
can be done without giving up stability of the model if
we add quartic potential terms that dominate at large
field values but are unimportant during preheating.
This model has ``negative coupling resonance", a mechanism
that was proposed in \cite{Tomislav}. Similar negative
coupling instabilities can also occur in models with cubic
interaction terms \cite{Natalia,Dufaux,Hassan}.

Another simple model in which the effective frequency
is negative for a certain time interval is the symmetry
breaking potential (\ref{pot1}). For small field values,
the effective mass of the fluctuations of $\phi$ is negative
and hence a ``tachyonic" resonance will occur, as studied
in \cite{Felder1,Felder2}. For small field values, the
equation for the fluctuations $\phi_k$ of $\phi$ is
\be
\ddot{\phi_k} + \bigl( k^2 - m^2 \bigr) \phi_k \, = 0 \, .
\ee
Hence, modes with $k < m$ grow with an exponent 
which approaches $\mu_k = 1$ in the limit $k \rightarrow 0$.
Given initial vacuum amplitudes for the modes $\phi_k$
at the intial time $t = 0$ of the resonance, the field
dispersion at a later time $t$ will be given by
\be
\langle \delta \phi^2 \rangle \, = \, \int_0^m \frac{k dk}{4 \pi} e^{2 t \sqrt{m^2 - k^2}} \, .
\ee
The growth of the fluctuations modes terminates once the
dispersion becomes comparable to the symmetry breaking
scale.

Tachyonic preheating also occurs in hybrid inflation models
like that of (\ref{pot2}). In this case, it is the fluctuations of
$\psi$ which have tachyonic form and which grow exponentially
\cite{Felder1}. Note that reheating in hybrid inflation was
first studied in \cite{Juan1} using the tools of broad
parametric resonance. Fermion production in this
context was discussed in \cite{Juan2}. Fermion production
is the context of tachyonic preheating was then analyzed
in \cite{Juan3}. The quantum to classical transition of
fluctuations in tachyonic preheating was investigated in
\cite{Juan4,Tranberg1}.

Another preheating mechanism which is more effective than
the broad resonance process described above arises if
the $\chi$ particles in the model of (\ref{toy}) are coupled linearly
to fermions such that the $\chi$ particles created when
$\phi \sim 0$ decay after half a $\phi$ oscillation, thus
preventing the $\chi$ particles from slowing the decay
of $\phi$. This mechanism is called ``instant preheating"
\cite{Felder3}.

\section{Thermalization}\label{section_thermalization}

\subsection{Perturbative Considerations}

Neither the perturbative decay of a Bose inflaton condensate
(discussed in Section 3.a) nor the preheating mechanism
discussed in later subsections of Section 3 produce a thermal
spectrum of decay products. For many questions in cosmology
it is not sufficient to know when inflation has terminated - it
is crucial to know at what temperature the universe first takes
on a thermal distribution. Examples are applications of reheating
to baryogenesis and to nucleosynthesis constraints. In this section
we first discuss perturbative thermalization. Then we summarize
the results of recent non-perturbative studies of inflaton decay.
We begin with perturbative considerations.

In a full thermal equilibrium the energy density
$\rho$ and the number density $n$ of relativistic particles
scale as: $\rho \sim T^4$ and $n \sim T^3$,
where $T$ is the temperature of the thermal bath. Thus, in
full equilibrium the average particle energy is given by:
$\langle E \rangle_{\rm eq} = \left( \rho/n \right)$, which obeys the scaling,
$\langle E \rangle_{\rm eq} \, \sim \, \rho^{1/4} \, \sim \, T $.

On the other hand, if the inflaton decays perturbatively,
then right after the inflaton decay has completed, the energy density
of the universe is given by:
\be \label{rho}
\rho \, \approx \, 3 \left(\Gamma m_{pl}\right)^2\,,~~~~~~
{\rm and}~~~~
\langle E \rangle \, \approx \, m \, \gg \, \rho^{1/4} \, ,
\ee
(where $m$ is the inflaton mass). Then, from
conservation of energy, the number density of decay products is found to be
\be \label{full}
n \, \approx \, \left({\rho \over m} \right) \, \ll \, \rho^{3/4} \, .
\ee
Hence, perturbative inflaton decay results in a dilute plasma that contains a
small number of very energetic particles.

Reaching full equilibrium requires re-distribution of the energy among
different particles, {\it kinetic equilibrium}, as well as increasing
the total number of particles, {\it chemical equilibrium}. 
Therefore both number-conserving and number-violating reactions must be involved.

The most important processes for kinetic equilibration
are $2 \rightarrow 2$ scatterings with gauge boson exchange
in the $t$-channel. (Scalar exchange in $t$-channel diagrams are usually
suppressed, also vertices that arise from a
Yukawa coupling are helicity suppressed.) The cross-section for these 
scatterings is
$\sigma_{2 \rightarrow 2} \sim \alpha^2 \vert t \vert^{-1}$,
where $\alpha$ is a gauge fine structure constant and the variable $t$ is
related to the exchanged energy $\Delta E$ and momentum,
$\Delta {\bf{p}}$ through
$t = {\Delta E}^2 - {\vert \Delta {\bf{p}} \vert}^2$.
Due to an infrared singularity, these scatterings are very efficient even in a dilute
plasma~\cite{Davidson:2000er,Allahverdi:2005mz}.

Chemical equilibrium is achieved by changing
the number of particles in the reheat plasma.
From (\ref{full}) it follows that
in order to reach full equilibrium the total number of particles
must {\it increase} by a factor of $n_{\rm eq}/n$, where $n \approx
\rho/m$ and the equilibrium value is: $n_{\rm eq} \sim
\rho^{3/4}$. This can be a very large number, e.g.
$n_{\rm eq}/n\sim {\cal O}(10^3)$.
It was recognized in~\cite{Davidson:2000er,Allahverdi:2002pu} (see
also~\cite{Jaikumar:2002iq,Allahverdi:2000ss}) that the most
relevant processes are $2 \rightarrow 3$ scatterings with gauge-boson
exchange in the $t-$channel.
The cross-section for
emitting a gauge boson whose energy is 
$E \sim \left(\Gamma m_{pl} \right)^{1/2}$
(where as in earlier sections of this review $\Gamma$ is the inflaton decay rate), 
from the scattering of two fermions (up to a logarithmic ``bremsstrahlung'' factor) is
$\sigma_{2 \rightarrow 3} \sim \alpha^3 \left(\Gamma m_{pl}\right)^{-1}$.
When these scattering become efficient, the number of particles
increases very rapidly~\cite{Enqvist:1993fm}
\footnote{Decay processes which were considered in~\cite{Allahverdi:2000ss}, are 
helpful, but in general they cannot increase the number of particles to the required level.} 
As a result, full thermal equilibrium will be established shortly after that.

Based on the above analysis, one can use the rate for the above
inelastic scatterings as the thermalization rate $\Gamma_{\rm th}$ of the universe . 
This rate at the time when the inflaton decay completes can be found by using 
Eqs.~(\ref{rho},\ref{full}):
\be \label{thermalization}
\Gamma_{\rm th} \sim \alpha^3 \left({m_{pl} \over m}\right) \Gamma .
\ee
For typical values of $\alpha \sim 10^{-2}-10^{-1}$ and $m \leq 10^{-5} m_{pl}$, we find $\Gamma_{\rm th} \geq \Gamma$. Therefore the universe reaches full thermal equilibrium immediately after the completion of perturbative inflaton decay.

\subsection{Non-Perturbative Considerations}

The purely perturbative considerations of the previous subsection are
subject to the same criticisms as the original perturbative analysis
of the initial stages of reheating. Hence, we must turn to non-perturbative
analyses. Some analytical approaches were pioneered in \cite{BHdV}
and \cite{Baacke}. Numerical studies, however, have
proved more powerful.

Since the occupation numbers of the excited
modes are typically very high after the initial stages of preheating,
a classical field theory analysis should be justified. Initial numerical
studies were pioneered in \cite{KT1,KT2,KT3,Tom,Felder4}. Two
numerical package to perform such simulations are publicly
available \cite{Easy,Defrost}. Detailed numerical simulations
of tachyonic preheating are given in \cite{Felder2}. Here, we will
focus on numerical studies of preheating in models with
narrow or broad parametric resonance \cite{Micha}.

The resonant phase of the reheating process described in Section 3
produces either (in the case of narrow resonance) field
fluctuations in a narrow interval about the resonant frequency
(which is set by the mass of the inflaton field), or else (in the
case of broad or tachyonic resonance) field fluctuations at
all wavenumbers smaller than the critical value given in
(\ref{ineq_adia2}), whose magnitude is set by the inflaton mass and
amplitude.

Once the occupation numbers of the resonant modes become
sufficiently large, re-scattering of the fluctuations begins. As
first studied in \cite{KT1,KT2,KT3,Tom} this terminates the
phase of exponential growth of the occupation numbers. In
the case of narrow resonance, new peaks in the spectrum of
the number density $n(k)$ develop at harmonic frequencies.
Soon, the spectrum shows excitations in a continuum band
which reaches to $k = 0$ and has an ultraviolet (UV) cutoff
whose value increases with increasing time.

As studied in detail in \cite{Micha}, the evolution of the field
fluctuations evolves to a regime of turbulent scaling driven
by the remnant oscillations of the inflaton condensate. The
resulting distribution of fluctuations is characterized by the
spectrum
\be
n(k) \, \sim \, k^{-3/2} \,
\ee
which is non-thermal (for a thermal distribution we would
have $n(k ) \sim k^{-1}$).

The evolution during the phase of turbulence has been
shown \cite{Micha} to be self-similar in the sense that as
a function of time the spectrum scales as
\be \label{scaling}
n(k, \tau) \, = \, \tau^{-q} n_0(k \tau^{-p}) \, ,
\ee
where $\tau$ is a rescaled time ($\tau = t / t_0$, where $t_0$
is the time when the turbulent scaling regime begins), $n_0$
gives the initial distribution of the particles, and $p$ and $q$ are
positive rational numbers whose values are determined in numerical
simulations. This equation (\ref{scaling}) describes the overall
growth in the number of fluctuation quanta and at the same time
gives the increase of the UV cutoff frequency as a function of time.

The phase of turbulence ends once most of the energy has
been drained from the inflaton field. At this time quantum
processes take over and lead to the thermalization of the
spectrum. In the case of an ${\cal O}(N)$ scalar field model,
the preheating and thermalization was considered analytically
in the large N approximation \cite{Serreau}, and applied to
thermalization of fermions and gauge fields
in hybrid inflation in \cite{Serreau2}
(see also \cite{Tranberg2} for a more general study of
thermalization of quantum fields in an expanding universe).

Note that the time interval of the resonant preheating phase
is given by the inflaton mass $m$ and hence is very short on
the Hubble time scale. The period when the intial
re-scatterings take place and which ends when the
turbulent scaling distribution becomes established is
longer than the period of the intial resonance, but
not by a large factor \cite{Micha}. Thus, the time $t_0$ is
of the same order of magnitude as the time when
inflation ends. However, the period of driven turbulence
is very long, in particular if all of the coupling constants
in the field theory model are small. If $c_r$ is the
fraction of the energy density $\rho_I$ at the end of inflation
which remains in the inflaton at the beginning of the
phase of turbulence, then a rough estimate of the time
$\tau_{\rm th}$ of thermalization is \cite{Micha}
\be
\tau_{\rm th} \, \sim \, \left( \frac{(c_r \rho_I)^{1/4}}{m} \right)^{1/p} \, .
\ee
A typical value of $p$ is $p = 1/7$. Given the normalization of $m$ 
from the observed magnitude of the cosmological
fluctuations we find $\tau_{\rm th} \sim c_r^{7/4} 10^{21}$. Hence, we 
see that the temperature at which thermal
equilibrium finally becomes established is low. The
resulting value of the reheating temperature in fact agrees roughly with 
what is obtained from the perturbative arguments given in
the previous subsection.

\section{Reheating in Supersymmetric Models}

As an important application of the theory of reheating described
in the previous sections we consider reheating in supersymmetric
models. Supersymmetry (SUSY) introduces new degrees of freedom and
new parameters, and a large number of scalar fields that may acquire large 
VEVs during inflation. These elements can affect various aspects of reheating 
that we discussed in Sections \ref{section_reheating} and \ref{section_thermalization}.
Here we demonstrate some of the possible effects in the context of a well-motivated 
SUSY model.

\subsection{Inflaton Couplings to Matter Fields}

The minimal supersymmetric SM (MSSM) is a well-motivated extension of the 
SM (for reviews see e.g.~\cite{Nilles:1983ge,Haber:1984rc}). The new fields in 
the MSSM are scalar partners of leptons and quarks, called sleptons and squarks 
respectively, and fermionic partners of gauge and Higgs fields called gauginos 
and Higgsinos respectively. The MSSM superpotential is given by
\begin{equation}
\label{mssm}
W_{\rm MSSM}=h_u { Q} { H}_u { u} + h_d { Q} { H}_d
{ d} + h_e { L} { H}_d { e}~ + \mu { H}_u { H}_d\,,
\end{equation}
where ${ H}_u, { H}_d, { Q}, { L}, { u}, { d},
{ e}$ in Eq.~(\ref{mssm}) are chiral superfields representing the
two Higgs fields (and their Higgsino partners), left-handed (LH)
(s)quark doublets, right-handed (RH) up- and down-type (s)quarks, LH
(s)lepton doublets and RH (s)leptons respectively. The
dimensionless Yukawa couplings $h_{u}, h_{d},
h_{e}$ are $3\times 3$ matrices in the flavor space, and we
have omitted the gauge and flavor indices. The last term is the
$\mu$ term, which is a SUSY version of the SM Higgs boson
mass.

Now we consider inflaton couplings to the MSSM fields. For a gauge singlet inflaton, 
the only renormalizable coupling occurs through the superpotential term 
$2 g { \Phi} { H_u} { H_d}$, with ${ \Phi}$ being the inflaton superfield 
(for details, see~\cite{Allahverdi:2007zz}). Taking into account 
the inflaton superpotential mass term $\left(m/2 \right){ \Phi} { \Phi}$,
the {\it renormalizable part of the
scalar potential} that is relevant for the inflaton decay into MSSM scalars is given by:
\begin{eqnarray}
\label{infpot}
V \supset {1 \over 2} m^2 {\phi}^2 + g^2 {\phi}^2 \chi^2_1 + g^2 \phi^2 \chi^2_2
+ {1 \over \sqrt{2}} g m \phi \chi^2_1 - {1 \over \sqrt{2}} g m \phi \chi^2_2 \,,
\end{eqnarray}
where $\chi_{1,2}$ denotes the scalar component of $(H_u \pm H_d)/\sqrt{2}$
superfields, and we have only considered the real parts of the
inflaton, $\phi$, and $\chi_{1,2}$ fields.

We see that even in the simplest SUSY set up, the
scalar potential is more involved than the non-SUSY case given in
Eq.~(\ref{intLag}), which in turn can alter the picture of preheating
presented in Section \ref{section_reheating} (see the detailed 
discussion in Refs.~\cite{Allahverdi:2007zz,Allahverdi:2005mz}). 
An interesting feature of Eq.~(\ref{infpot}) is that both the cubic 
$\phi \chi^2$ and quartic $\phi^2 \chi^2$ interactions appear and SUSY naturally
relates their strengths \footnote{Note that the cubic term is required for a complete decay of 
the inflaton field.}. Also, the inflaton coupling to fermionic partners of $\chi_{1,2}$ follows 
naturally  from SUSY. The prospects for fermionic preheating will thus be the same as 
those for the bosonic case.

\subsection{Supersymmetric Flat Directions}

A key property of SUSY theories is the presence of flat directions in 
field space along which the potential identically vanishes (in the limit of unbroken SUSY). Such 
scalar fields (which are complex) can therefore obtain large VEVs along these special directions at no 
energy cost. These flat directions, which can be interpreted as a degeneracy of the vacuum 
state of SUSY theories, arise because of cancellations between fields of opposite charges 
in the D-term potential. A powerful tool for finding the flat directions
has been developed in \cite{Dine:1995uk,Dine:1995kz,Gherghetta:1995dv} (for  reviews 
see~\cite{Enqvist:2003gh,Dine:2003ax,Basboll:2009tz }). Flat directions are classified by 
gauge-invariant monomials $\prod^{n}_{i=1} { X_i}$, where $X_i$ are chiral 
superfields of the model. This ensures that the $D$-term part of the potential 
vanishes\footnote{Since the total SM charge of a gauge-invariant monomial is zero by 
definition, the D-term potential involving only the fields used to build the monomial will 
also vanish since it is proportional to the sum of the charges.} 
along the direction $\langle \chi_1 \rangle = ... = \langle \chi_n \rangle = \varphi$ ($\chi_i$ 
are scalar components of $X_i$). This corresponds to a two-dimensional subspace 
represented by a complex field $\varphi$. A flat direction VEV spontaneously breaks gauge 
symmetries and gives (SUSY conserving) masses to the gauge bosons/gauginos similar to 
the Higgs mechanism in electroweak symmetry breaking
~\cite{Dine:1995kz,Allahverdi:2005mz,Allahverdi:2007zz,Allahverdi:2008pf,Allahverdi:2006xh}. 
The induced masses for gauge/gaugino fields are $\sim \alpha^{1/2} \vert \varphi \vert$ 
(we recall that $\alpha$ is a gauge fine structure constant). Similarly, a flat direction VEV 
induces (SUSY conserving) masses $\sim h \vert \varphi \vert$ for those fields 
that have superpotential couplings to $\varphi$ ($h$ is a Yukawa coupling). 
Therefore all fields that are coupled to a flat direction obtain very large masses.

The flat directions are massless if SUSY is exact, but they are lifted when SUSY is broken
(which is assumed to happen at a scale of the order of TeV), as a result of 
which they get a mass $m_\varphi \sim {\cal O}({\rm TeV})$. Provided that 
$m_\varphi \ll H_{\rm inf}$, $H_{\rm inf}$ being the Hubble expansion rate during inflation, 
the flat direction can acquire a large VEV by the virtue of quantum jumps during inflation, 
see the discussion in Refs.~\cite{Enqvist:2003gh,Dine:2003ax}. This can dramatically 
alter the post-inflationary history of the universe as we will see in the next subsections 
\footnote{The development of large VEVs requires that the flat directions do not obtain 
positive Hubble-induced supergravity corrections during inflation. This problem can be 
avoided, for example, by considering non-minimal Kahler potentials.}.

\subsection{Perturbative Decay}

Consider a flat direction $\varphi$ that has Yukawa couplings to the inflaton decay 
products $\chi$. This happens, for example, for MSSM flat directions that are made of 
squark and/or slepton fields with $\chi$ being a MSSM Higgs field, see 
Eq.~(\ref{infpot}) (for details, see~\cite{Allahverdi:2007zz}). This results in the following 
term in the scalar potential:
\begin{equation} \label{chimass}
V \, \supset \, h^2 {\vert \varphi \vert}^2 {\chi^2} ,
\end{equation}
where $h$ denotes a Yukawa coupling.
Note that the first generation of leptons and quarks have a Yukawa coupling 
$\sim {\cal O} (10^{-5})$, while the rest of the SM Yukawa couplings are $> 10^{-4}$. 
Since $\vert \varphi \vert$ is virtually frozen while $m_{\varphi} < H < H_{\rm inf}$ 
it is only when $H \simeq m_{\varphi}$ that the flat direction starts its 
oscillations. Since the field is complex, typically an elliptical trajectory with an 
${\cal O}(1)$ eccentricity will result \cite{Dine:1995kz}. Hence, $\vert \varphi \vert$ 
will  redshift as  $\vert \varphi \vert \propto H^{-1}$. 

While the flat direction has
a large amplitude, the induced mass of the inflaton decay products obtained via
(\ref{chimass}) will lead to the inflaton decay being kinematically forbidden as 
long as $h \vert \varphi \vert \geq m/2$. 
There are thus two criteria for perturbative inflation decay. First of all, the
decay of the inflaton into $\chi$ particles must be kinematically allowed
which will become possible once the induced $\chi$ mass drops below
the inflaton mass $m$. Taking into account the fact that once $H$ falls
below the value $m_{\varphi}$  the field amplitude of $\varphi$ decreases
linearly in $H$ we find that the kinematic decay becomes possible once
$H < \left({m \over h \varphi_0} \right) m_\varphi$, 
where $\varphi_0$ is the initial VEV of the flat direction. A second condition
for perturbative inflaton decay to occur is that $H < \Gamma$,  
$\Gamma$ being the rate for perturbative decay of the inflaton. Thus,
the inflaton cannot decay until the Hubble rate has decreased to a value $H_{\rm dec}$
given by:
%
\begin{equation} \label{dec1}
H_{\rm dec} \, = \, {\rm min} ~ \left[\left({m \over h \varphi_0} \right) m_\varphi ,~~
\Gamma \right]\, .
\end{equation}
If $\varphi_0$ is sufficiently large, then we can have $H_{\rm dec} \ll \Gamma$. This happens if
%
$\varphi_0 \, > \, h^{-1} {m_\varphi m \over \Gamma} $.
%
Flat directions can therefore significantly delay inflaton decay on purely kinematical grounds.

\subsection{Non-perturbative Decay}

In order to understand the preheating dynamics in the presence of flat directions, we consider the
governing potential that is obtained from Eqs.~(\ref{infpot},\ref{chimass}):
\begin{equation} \label{mssm1}
V \, = \, {1 \over 2} m^2 {\phi}^2 + g^ 2 {\phi}^2 {\chi}^2 +
{g \over \sqrt{2}} m \phi {\chi}^2 + h^2 {\vert \varphi \vert}^2 {\chi}^2.
\end{equation}
As mentioned in the previous section, we generically have $h > 10^{-4}$, and 
$g$ can be as large as $\sim {\cal O}(1)$.
After mode decomposition of the field $\chi$, the energy of the mode
with momentum $k$, denoted by $\chi_{k}$, is given by:
\begin{equation} \label{energy}
\omega_k \, = \, {\left(k^2 + 2 g^2 \phi^2 +
\sqrt{2} g  m \phi +
2 h^2 {\vert \varphi \vert}^2 \right)}^{1/2}.
\end{equation}
Let us freeze the expansion of the universe first. Including the expansion will not 
change our conclusions.

We will now show that there is also a kinematic blocking of preheating if the
initial value of the flat direction fields is large.
We consider the most efficient case for preheating, large field inflation, e.g.
$\langle \phi \rangle > m_{pl}$. Note that for $g > 10^{-6}$, the inflaton induces a large
mass $g \langle \phi \rangle > H_{\rm inf}$ for $\chi$ during inflation.
As a result, $\chi$, quickly settles down to the minimum even if it is initially displaced, and
remains there. Therefore, $\varphi$, does not receive any mass
corrections from its coupling to $\chi$ during inflation.

As discussed in the previous subsection, in the interval 
$m_\varphi \leq H \leq m$, the flat direction VEV slides very 
slowly because of the under damped motion due to
large Hubble friction term - it is effectively frozen.
Non-perturbative production of $\chi$ quanta will occur if
there is a non-adiabatic time-variation in the energy, e.g. that ${d
\omega_k/dt} \geq \omega^2_k$. The inflaton oscillations result in a
time-varying contribution to $\omega_k$, while the flat direction
coupling to $\chi$ yields a virtually {\it constant} piece. 
This constant piece weakens the non-adiabaticity condition. Indeed time-variation
of $\omega_k$ will be adiabatic at all times, i.e.
${d \omega_k / dt} < \omega_k^2$, provided that
$h^2 \vert \varphi \vert^2 > g \Phi m$,
where $\Phi \sim {\cal O}(m_{pl})$ is the amplitude of the inflaton oscillations.
Thus, there will be no resonant production of $\chi$ quanta if
\begin{equation} \label{cond}
\varphi_0 \,  >  \, h^{-1} \left(g m_{pl} m \right)^{1/2}\, ,
\end{equation}
Similar arguments lead to a kinematical blocking
of fermionic preheating, as the symmetry between bosons and fermions implies similar
equations for the momentum excitations, see Eq.~(\ref{energy}).

\subsection{Thermalization}

The flat direction VEV spontaneously breaks the SM gauge
group. The gauge fields of the broken symmetries then acquire a
SUSY conserving mass of the order of $\alpha^{1/2} \vert \varphi \vert$. This
mass provides a physical infrared cut-off for scattering diagrams with
gauge boson exchange in the $t-$channel. Thus the cross-section for 
inelastic scatterings is now given by 
$\sigma_{2 \rightarrow 3} \sim \alpha^2 \vert \varphi \vert^{-2}$. 
For large values of $\vert \varphi \vert$ the scattering rate is suppressed, 
which results in a delayed thermalization.
It can be shown that the universe reaches thermal equilibrium when the 
Hubble expansion rate is (for details see Ref.~\cite{Allahverdi:2005mz})
\ba \label{thermal1}
H_{\rm th} \, &=& \, {\rm min} ~ [10 \alpha^2 \left({m_{pl} \over \varphi_0}\right)^2 {m^2_\varphi \over m} , \\
& &~~ 10 \alpha^2 \left({m_{pl} \over \varphi_0}\right)^2 \left({\Gamma \over m_\varphi}\right)^{1/2} {m^2_\varphi \over m} ,~~ \Gamma ] , \nonumber
\ea
where $\Gamma$ is the rate for perturbative inflaton decay. This yields the
following expression for the reheat temperature 
\be \label{reheat1}
T_R \, \sim \, \left(H_{\rm th} m_{pl} \right)^{1/2} .
\ee

For very large values of $\varphi_0$, thermalization is considerably delayed, e.g. 
$H_{\rm th} \ll \Gamma$, and hence $T_R \ll \left(\Gamma m_{pl} \right)^{1/2}$. 
This happens for
\be
\varphi_0 \, > \, 3 \alpha {m_\varphi m_{pl} \over (m \Gamma)^{1/2}} .
\ee
%

In the above discussions it has been assumed that the flat direction condensate 
does not decay non-perturbatively. One may think that non-perturbative effects 
could also result in a fast decay of flat directions similar to preheating \cite{Natalia}. 
However, there is crucial difference between a rotating and a radially oscillating 
condensate. The $F$-term couplings do not lead to resonant particle production 
from a rotating condensate~\cite{Allahverdi:1999je,Postma:2003gc}. It has been 
shown that $D$-term couplings of the flat direction can result in non-perturbative 
particle production~\cite{Olive:2006uw}. However, unlike the case of a radially 
oscillating field, the produced quanta have momenta that are less than the mass 
of the condensate in this case.

The reason is that a rotating condensate has a $U(1)$ global 
charge\footnote{This charge corresponds to the angular momentum of the rotating 
condensate in field space.} that, in the case 
of MSSM flat directions, is identified with the baryon and/or lepton 
number~\cite{Affleck:1984fy,Dine:1995kz,Dine:1995uk}. Now, it is easy to show that the 
charge per particle in the rotating condensate is of $\mathcal{O}(1)$. Note also that the 
baryon and lepton 
number of MSSM fields, whether in the flat direction condensate or produced 
from the flat direction rotation, is $\pm 1/3$ and $\pm 1$ respectively. Hence preservation 
of the baryon/lepton number by the $D$-term interactions, along with the conservation 
of energy density, implies that for an elliptic trajectory with ${\cal O}(1)$ eccentricity 
the number of produced quanta cannot be much smaller than the number of 
zero-mode quanta in the rotating condensate~\cite{Allahverdi:2008pf}. Therefore, the 
non-perturbative decay of a rotating flat direction does not change the thermalization 
picture described above. 

The reason for a delayed thermalization is due to inducing large masses to the 
gauge/gaugino fields from the VEV of the flat direction. 
The induced mass by the plasma is given by 
$m^2_{eff} \sim \alpha n/\langle E \rangle$, where $\langle E \rangle$ and $n$ are 
the average energy and number density of quanta in the plasma respectively. Since they 
cannot be much different from those in the condensate, the induced masses will also 
be comparable.

\section{Consequences of Reheating/Preheating}

\subsection{Non-thermal Particle Creation}

Reheating and preheating lead to non-thermal particle production, as we
have seen in previous sections. In cosmology it is usually assumed that
all particles start out in thermal equilibrium at the beginning of the
Standard Cosmology phase. However, reheating begins with out-of-equilibrium
decay of the inflaton oscillations and, as we discussed, decay products may not reach full thermal equilibrium immediately. During the transition from inflation to the Standard Cosmology various non-thermal processes take place and the assumption of thermal equilibrium of all particles clearly breaks down. In the following, we briefly mention a few applications of non-thermal particle production.

\subsubsection{Baryogenesis and Leptogenesis:}

The first application is to baryo- and leptogenesis. One of the
several possible mechanisms to explain the observed asymmetry
between baryons and antibaryons is to make use of out-of-equilibrium
decay of superheavy Higgs and gauge particles~\cite{BG}. If
reheating were purely perturbative, particles as heavy as the inflaton could be
created either in inflaton decay~\cite{Allahverdi:2002nb} or from scatterings of inflaton decay products~\cite{WIMPZ,Allahverdi:2002pu}. 

Preheating, however, provides a mechanism to produce a large population
of superheavy scalar particles much heavier than the inflaton. In \cite{KLR} this was studied making
use of the same chaotic inflation Lagrangian which we have
used in Section 3, with the $\chi$ scalars being the superheavy
Higgs or gauge fields. A full numerical study \cite{KRT}
showed, however, that large self-interactions may
terminate the resonance before it becomes effective.

Another way to generate to observed baryon to entropy ratio is
via leptogenesis \cite{LG}, a scenario in which initially an asymmetry
in the lepton number is produced that is then partially converted 
into baryon asymmetry via SM sphalerons
\cite{KRS}. Preheating after inflation is a way to generate the
initial lepton asymmetry. For example \cite{GT}, preheating can produce
a large number density of super-massive RH neutrinos
in a model in which the inflaton couples to these neutrinos $\psi$
via the standard fermionic preheating interaction term
\be\label{fermionic-coupling}
{\cal L}_I \, = \, g \phi {\bar \psi}{\psi} \, .
\ee
If hybrid inflation occurs at a scale close to the electroweak
scale, then the non-thermal production of particles may
provide the out-of-equilibrium condition that is necessary
in order to achieve electroweak baryogenesis \cite{EWBG}.

\subsubsection{Dark matter:}

Another application of non-thermal particle creation during
reheating is to excite dark matter. It is usually assumed that
the dark matter particles are thermally
distributed. This assumption is implicit in most
current analyses of the prospects for dark matter
detection in direct and indirect experiments. However,
if the dark matter particles couple to the inflaton,
then non-thermal production of dark matter during
reheating is to be expected. If the dark matter
particles have sufficiently strong interactions which
allows them to
thermalize during reheating, then the signatures
of the initial non-thermal distribution will be washed
out. However, if the interactions do not permit
thermalization after inflation, then the predictions
concerning the dark matter distribution will be
quite different.

The production of out-of-equilibrium dark matter during
preheating was put forwards in \cite{GT} and then
studied in detail in \cite{WIMPZ}. In the latter
reference, the superheavy dark matter particles produced
during reheating were called ``Wimpzillas". Masses
of Wimpzillas comparable to the grand unified theory (GUT) scale were
considered (see also \cite{Kolb} for a discussion
of a purely gravitational production mechanism for
Wimpzillas). The dark matter abundance which can
be obtained by the preheating channel is very
model-dependent, whereas direct gravitational
particle production produces dark matter of the
required abundance for particle masses of
$M_X \sim g^{1/2} 10^{15}$~GeV~\cite{Kolb}.

\subsubsection{Moduli and Gravitino Production}

Preheating could also produce dangerous and
unwanted particles \cite{GTR1}. An example are particles with gravitationally 
suppressed couplings and weak scale masses that
arise in many theories beyond the SM. Overproduction of these particles 
could overclose the universe, if they are stable, or ruin the success of Big Bang 
nucleosynthesis (BBN) in the case of unstable relics. Here we consider moduli 
and gravitino production during preheating.

The existence of bosonic and fermionic moduli fields is common in SUSY and 
superstring theories.  Moduli  (bosonic modulus, $\chi$, and fermionic modulus, 
$\psi$) are typically coupled to the inflaton via
non-renormalizable interaction terms such as
\be
{\cal{ L}} \, \sim \, \phi^4 \frac{\chi}{m_{pl}}~~~~
({\rm bosonic}),~~~~~~~~~
{\cal{L}} \sim  \frac{\phi^2}{m_{pl}} {\bar{\psi}}\psi ~~~~({\rm fermionic})\,.
\ee
The production of moduli fields in chaotic and hybrid inflation reheating
was analyzed in \cite{GRK}. It was shown that moduli
field can be parametrically amplified (their amplitude
remains smaller than that of the inflaton fluctuations).

Another important example is the gravitino, the spin 3/2 partner of the graviton.
Gravitinos are produced thermally from scatterings of light particles in the thermal bath.
The number density of gravitinos thus produced 
can be obtained by solving the Boltzmann equation:
\be
\dot n_X + 3 H n_X \, \simeq \, \langle \sigma v \rangle n_l^2 \, ,
\ee
where $n_X$ is the number density of the gravitinos,
$\sigma$ is the production cross section which scales
as $m_{pl}^{-2}$, and $v \sim c$ is the relative velocity of scatterers $l$ 
whose number density is $n_l$. The
resulting abundance is found to be~\cite{Ellis,Buchmuller}
\be
\frac{n_X}{s} \, \sim \, 10^{-2} \frac{T_R}{m_{pl}} \, ,
\ee
where $s$ is the entropy density and $T_R$ is the reheat temperature of the universe.  
BBN gives rise to an absolute upper bound $(n_X/s)< 10^{-12}$ (the exact number 
depends on the gravitino mass and its decay modes), which in turn leads to an upper 
bound $T_R < 10^9$ GeV (see e.g.~\cite{Moroi,CEFO}).

Gravitino production during preheating was studied
in \cite{Maroto}. The gravitino
equation of motion is the Rarita-Schwinger equation.
Conformal invariance is broken during the reheating phase
of inflationary cosmology. The presence of the oscillating
inflaton field leads to a periodically varying correction
to the effective gravitino mass that results in an
instability in the same way that there is an instability for spin 0 and 1/2 particle
modes. The exact strength of the instability depends
sensitively on the precise SUSY inflationary
model one is considering. 

Gravitino with helicity $\pm 1/2$ component mainly contain the Goldstino 
component- the inflatino (superpartner of the inflaton),
whose interactions are not suppressed by $m_{pl}$. One would naturally 
expect them to be created in large abundance. However,
in realistic scenarios, where the scale of inflation is much higher than the 
scale of SUSY breaking, e.g.  $H_{inf}\gg 
{\cal O}(100~{\rm GeV})$, it was argued in~\cite{ABM} and explicitly shown 
in~\cite{NPS} that the helicity $\pm 1/2$ states that are produced during 
preheating mainly decay in the form of inflatinos along with the inflaton.  

\subsection{Metric Preheating}

\subsubsection{Entropy fluctuations}

As we have seen repeatedly in this article, the oscillating inflaton field has
potential to lead to preheating of any fields it couples to. The metric itself
is no exception. Metric fluctuations come in three types (for details see
a review article on the theory of cosmological perturbations, e.g. \cite{MFB})
- scalar modes, vector modes and tensor modes. In an expanding universe
the vector modes are negligible since they decay. The tensor modes
represent gravitational waves. The scalar modes are the ``cosmological
perturbations" which are sourced by matter fluctuations.

It is possible (see e.g. \cite{MFB}) to choose a coordinate system in which
the metric including scalar metric fluctuations is diagonal:
\be\label{metric}
ds^2 \, = \, a^2(\eta)\left[(1+2\Phi)d\eta^2-(1-2\Psi)\gamma_{ij}dx^idx^j\right], \,
\ee
where $\Phi$ and $\Psi$ are the two scalar metric fluctuation potentials.
They are functions of space and time. In the absence of anisotropic stress
(e.g. for scalar field matter) the two potentials are in fact equal. In the above,
$a(\eta)$ is the scale factor of the background cosmology and $\eta$ is
conformal time defined via $dt = a d\eta$. Also, $\gamma_{ij}$ is the
background metric of the spatial sections after factoring out the cosmological
expansion.

In the case of scalar field matter the equation of motion for the metric fluctuation
variable $\Phi$ reads (see e.g. \cite{MFB})
\be\label{mastereq}
\Phi'' + 2 \bigl( {\cH} - \frac{\phi_{0}''}{\phi_{0}'} \bigr)  {\Phi} -  \nabla^2 {\Phi}
+2  \bigl( {\cH}' - {\cH}\frac{\phi_0''}{\phi_0'} \bigr) {\Phi} \, = \, 0 \, ,
\ee
where a prime indicates the derivative with respect to $\eta$ and $\cH$ is the
Hubble expansion rate in conformal time. Also, $\phi_0$ denotes the
background value of the scalar field. Note that in the above we are
assuming that the perturbations are purely adiabatic (the relative density
fluctuations in each matter component are the same). If there are entropy
fluctuations present, there will be a source term on the right-hand side
of (\ref{mastereq}) which is proportional to the entropy fluctuation.

It appears from (\ref{mastereq}) that the oscillations of the background
inflaton field could induce parametric resonance of the metric fluctuations
\cite{Bassett1}. From our studies of preheating in Section 3 we would
expect the long wavelength modes to be the most sensitive to this
instability. However, a careful study \cite{Fabio1} (see also
\cite{Zhang,Niayesh}) showed that there
is no instability of adiabatic metric perturbations during reheating.
This can most easily be seen by focusing on a new variable $\zeta$,
the curvature fluctuation on constant density hypersurfaces, which is given
by
\be\label{def_zeta}
\zeta \, \equiv \, \frac{2}{3}\frac{H^{-1}\dot{\Phi}+\Phi}{1+w}+\Phi \, ,
\ee
where $w = p / \rho$ is the equation of state parameter.
In the case of adiabatic fluctuations, $\zeta$ on scales larger
than the Hubble radius $H^{-1}$ satisfies the equation
\be \label{zetaeq}
{\dot{\zeta}} (1 + w) \, = \, 0 \, .
\ee
In spite of
the fact that during the inflaton oscillations $w = -1$ is reached, it can
be shown that $\zeta$ does not change.

However, for entropy fluctuations the result is rather different
\cite{Bassett2, Fabio2}. Provided that the entropy field (e.g.
the $\chi$ field in the chaotic inflation preheating discussion
of Section 3 or the field $\psi$ in the hybrid inflation model)
undergoes a parametric or tachyonic instability and thus leads
to an exponential growth of the entropy fluctuation $\delta S$,
then the curvature fluctuation $\zeta$ will inherit this exponential
growth. This follows since the entropy fluctuations act as a
source for $\zeta$ - the equation (\ref{zetaeq}) gets replaced by
\be
{\dot{\zeta}} \, = \, \frac{\dot{p}}{p + \rho} \delta S \, .
\ee

This resonant growth of entropy fluctuations only is important
in models in which the entropy fluctuations are not suppressed
during inflation. Some recent examples were studied in
\cite{BFL,ABD}. The source for this instability need not be
the oscillations of the inflaton field. In SUSY
models, the decay of flat directions can also induce this
instability \cite{Francis}.

\subsubsection{Gravity waves:}

The equation of motion for gravitational waves is similar to that
of scalar metric fluctuations (\ref{mastereq}) except that there
is no coupling between the oscillating scalar field and the
wave amplitude. Hence, there is no direct preheating of
gravitational waves.

Nevertheless, gravitational waves can be produced by secondary
processes. As analyzed in \cite{KT4}, gravitational waves can
be produced from the interaction of the classical matter waves
produced during preheating. This is a re-scattering effect. The
induced gravitational wave spectrum is not scale-invariant but
has a pronounced peak whose frequency is determined by
the scale of inflation. For an inflation energy scale of $10^{15} \rm{GeV}$,
the peak of the spectrum is at about $10^8 \rm{Hz}$. In
hybrid inflation models, the scale of the peak can be in the $\rm{kHz}$
range relevant for Advanced LIGO \cite{GB0}. The amplitude of the
peak is a couple of orders of magnitude higher than the
scale-invariant background of gravity waves produced
directly during inflation.

Recently, several groups \cite{Easther,GB,Dufaux2,Siemens}
have performed improved analyses of gravitational wave
production during preheating. The formalism of
\cite{Easther,Dufaux2,Siemens} are quite general and were
applied e.g. to a $\lambda \phi^4$ model of inflation
\cite{Siemens}. In \cite{Dufaux3}, the formalism for the
generation of gravitational waves was applied to hybrid
inflation, and \cite{GB} considered mostly gravitational
wave production by the collision of bubbles formed during
the tachyonic resonance in hybrid inflation models. The
tachyonic instability leads to the formation of bubbles,
and the collision of the bubbles is the primary source
of gravitational waves. In presence of fermionic couplings, see 
Eq.~\ref{fermionic-coupling}, the inflaton can fragment to form
non-topological solitons~\cite{KSM}, since the fragmentation 
of the inflaton condensate is inhomogeneous and anisotropic, 
it leads to large production of gravity waves as shown in~\cite{KMM}.

\section{Discussion and Conclusions}

We have presented an overview of theory and application of reheating
in inflationary cosmology. Particular emphasis has been on the
preheating mechanisms which in many models leads to rapid energy transfer
between the inflaton and regular matter.

Our discussion of
applications of reheating has been superficial due to lack of space.
We have in fact not discussed a number of issues such as topological
defect production during preheating \cite{KL1,KL2,Parry}, magnetic
field generation \cite{mag1,mag5},
induced non-Gaussianities
(see e.g. \cite{Anupam1,Barnaby1,Kohri,Rajantie1}),
preheating in theories with non-Standard kinetic terms (see e.g. \cite{Lachap} or extra
dimensions (possible excitation of Kaluza-Klein modes), applications to multi-field
inflation models \cite{Diana} and
to reheating in brane inflation models \cite{braneinfl}, and
effects of noise on reheating \cite{Craig1,Craig2} (leading to a new
proof \cite{Craig3} of Anderson localization).

It is important to point out that resonant phenomena which are important
in reheating can also play a role in other areas of cosmology where there
are oscillating scalar fields. One example is in the context of the MSSM
where oscillating moduli fields can lead to resonances 
\cite{Allahverdi:2006iq,Francis}. The
study of resonant effects in early universe cosmology is a rich area of
research which has of now barely been touched. Many of the lessons learned
in the context of inflationary reheating have more general applicability.

\begin{acknowledgments} 
 
This work is supported in part (RB) by a NSERC Discovery Grant, and by funds from
the Canada Research Chairs Program. 
RB also acknowledges support from a Killam Research 
Fellowship. FYCR is recipient of a NSERC CGS D scholarship. AM is partly
supported by the Marie Curie Research and Training Network \#8220, UniverseNet \#8221,
MRTN-CT-2006-035863. We wish to thank Juan Garcia-Bellido, J. Serreau, A. Tranberg 
and M. Trodden for comments on the draft.

\end{acknowledgments} 


\end{document}